\documentclass[aps,prl,twocolumn,pacs]{revtex4}
\usepackage{amsmath,bm,epsfig}
\usepackage{epsfig}
\usepackage{bm}
\newcommand{\B}[1]{{\bm{#1}}}
\newcommand{\C}[1]{{\mathcal{#1}}}
\newcommand{\pa}{\partial}
\usepackage[latin1]{inputenc}

\begin{document}
\title{Weakly Nonlinear Theory of Dynamic Fracture}
\author{Eran Bouchbinder, Ariel Livne and Jay Fineberg}
\affiliation{Racah Institute of Physics, Hebrew University of Jerusalem, Jerusalem 91904, Israel}

\begin{abstract}
The common approach to crack dynamics, linear elastic fracture
mechanics (LEFM), assumes infinitesimal strains and predicts a
$r^{-1/2}$ strain divergence at a crack tip. We extend this
framework by deriving a weakly nonlinear
fracture mechanics theory incorporating the leading nonlinear elastic
corrections that must occur at high strains. This yields strain
contributions ``more-divergent'' than $r^{-1/2}$ at a finite distance from the tip and
logarithmic corrections to the parabolic crack tip opening
displacement. In addition, a dynamic length-scale, associated with
the nonlinear elastic zone, emerges naturally. The theory
provides excellent agreement with recent near-tip measurements that can not be described in the LEFM framework.
\end{abstract}
\maketitle

Understanding the dynamics of rapid cracks is a major challenge in
condensed matter physics. For example, high velocity crack tip
instabilities \cite{99MF, 07LBDF} remain poorly understood from a
fundamental point of view. Much of our understanding of how
materials fail stems from Linear Elastic Fracture Mechanics (LEFM)
\cite{98Fre}, which assumes that materials are {\em linearly}
elastic outside of a small zone where all
nonlinear and dissipative processes occur. A central facet of LEFM
is that strains diverge as $r^{-1/2}$ at a crack's tip and that this
singularity dominates all other strain contributions in this region.
Linear elasticity should be expected to break down before
dissipative processes occur. The small size and rapid propagation
velocity of the near-tip region of brittle cracks have, however,
rendered quantitative measurements of the near-tip fields elusive.

In the companion Letter \cite{companion} such direct near-tip
measurements of the displacement field $\B u(\B r)$ were achieved
for Mode I cracks propagating at rapid velocities, $v$. Defining
$(r,\theta)$ as coordinates moving with the crack tip, the
propagation direction, $x$ is defined by $\theta\!\!=\!\!0$ and the
loading direction, $y$, by $\theta \!=\!\pi/2$. As predicted by
LEFM, these experiments revealed that the crack tip opening profile,
$u_y(r,\pm\pi)$, is parabolic beyond a velocity-dependent
length-scale $\delta(v)$. However, it was shown that although
$u_x(r,\theta\!=\!0)$ in this range also follows the functional form
predicted by LEFM, its parameters are {\em inconsistent} with those
described by $u_y(r,\pm\pi)$! Moreover, the strain component
$\varepsilon_{yy}(r,0)=\pa_yu_y(r,0)$ was {\em wholly incompatible}
with LEFM, indicating a ``more-divergent'' behavior than $r^{-1/2}$.
These puzzling discrepancies become increasingly severe as $v$
increases.

In this Letter, we show that {\em all} of these puzzles can be
quantitatively resolved by taking into account nonlinear corrections
to linear elasticity, which {\em must} be relevant near the crack
tip. This is achieved by perturbatively expanding the momentum balance equation for an elastic medium up to second order nonlinearities in the displacement gradients. The resulting theory provides a novel picture of the structure
of the fields surrounding a crack tip, and may have implications for
our understanding of crack dynamics.

Nonlinear material response at the large strains near a crack's tip
motivates us to formulate a nonlinear elastic dynamic fracture
problem under plane stress conditions. Consider the deformation
field $\B \phi$, which is assumed to be a continuous, differentiable
and invertible mapping between a reference configuration $\B x$ and
a deformed configuration $\B x'$ such that $\B x'\!=\!\B \phi(\B
x)=\B x+\B u(\B x)$. The deformation gradient tensor $\B F$ is
defined as $ \B F\!=\!\nabla \B \phi$ or explicitly
$F_{ij}\!=\!\delta_{ij}+\pa_j u_i$. The first Piola-Kirchhoff stress
tensor $\B s$, that is work-conjugate to the deformation gradient
$\B F$, is given as $\B s\!=\!\pa_{\B F} U(\B F)$, where $U(\B F)$
is the strain energy in the deformed configuration per unit volume
in the reference configuration \cite{Holzapfel}. The momentum balance equation is
\begin{equation}
\label{EOM}
\nabla \cdot \B s = \rho \pa_{tt}{\B \phi} \ ,
\end{equation}
where $\rho$ is the mass density. Under steady-state propagation
conditions we expect all of the fields to depend on $x$ and $t$
through the combination $x\!-\!vt$ and therefore
$\pa_t\!=\!-v\pa_x$. The polar coordinate system that moves with the
crack tip is related to the rest frame by
$r\!=\!\sqrt{(x-vt)^2+y^2}$ and $\theta\!=\!\tan^{-1}[y/(x-vt)]$.
Thus, the traction-free boundary conditions on the crack faces are
\begin{equation}
\label{BC}
s_{xy}(r,\theta\!=\!\pm\pi)\!=\!s_{yy}(r,\theta\!=\!\pm\pi)=0 \ .
\end{equation}

To proceed, we note that in the measurement region of
\cite{companion} the maximal strain levels are $0.2\!-\!0.35$ (see
below) as the velocity of propagation varied from $0.20c_s$ to
$0.78c_s$, where $c_s\!=\!\sqrt{\mu/\rho}$ is the shear wave speed
($\mu$ is the shear modulus). These levels of strain motivate a
perturbative approach where quadratic elastic nonlinearities must be
taken into account. Higher order nonlinearities are neglected below,
though they most probably become relevant as the crack velocity
increases. We write the displacement field as
\begin{equation}
\label{expansion}
\B u(r,\theta) \simeq \epsilon \B u^{(1)}(r,\theta)+\epsilon^2 \B u^{(2)}(r,\theta)+ \C O(\epsilon^3) \ ,
\end{equation}
where $\epsilon$ quantifies the (dimensionless) magnitude of the
strain. For a general $U(\B F)$, $\B s$ and $\B \phi$ can be
expressed in terms of $\B u$ of Eq. (\ref{expansion}). Substituting
these in Eqs. (\ref{EOM})-(\ref{BC}) one can perform a controlled
expansion in orders of $\epsilon$.  

To make the derivation concrete, we need an explicit $\B U(\B F)$ that corresponds
to the experiments of \cite{companion}. The polymer gel used in
these experiments is well-described by a plane stress incompressible
Neo-Hookean constitutive law \cite{05LCF}, defined by the energy
functional \cite{83KS}
\begin{equation}
\label{NH} U(\B F)= \frac{\mu}{2}\left[ F_{ij}F_{ij}+\det(\B
F)^{-2}-3\right] \ .
\end{equation}

Using this explicit $\B U(\B F)$, we derive the first order problem in $\epsilon$
\begin{equation}
\mu\nabla^2{\B u^{(1)}}+3\mu\nabla(\nabla\cdot{\B u^{(1)}})=\rho\ddot{\B u}^{(1)} \ ,
\label{Lame}
\end{equation}
with the boundary conditions at $\theta\!=\!\pm\pi$
\begin{equation}
\label{BC1}
r^{-1}\pa_\theta u_x^{(1)}+\pa_r u_y^{(1)}=0,\quad 4r^{-1}\pa_\theta u_y^{(1)}+2\pa_r u_x^{(1)}=0 \ .
\end{equation}
This is a standard LEFM problem \cite{98Fre}. The near crack-tip
(asymptotic) expansion of the steady state solution for Mode I
symmetry is \cite{98Fre}
\begin{eqnarray}
\epsilon u_x^{(1)}(r, \theta;v)\!&=&\!\frac{K_I \sqrt{r}}{4\mu\sqrt{2\pi}}\Omega_x(\theta;v)\!+\!\frac{Tr\cos\theta}{3\mu}+\C O(r^{3/2}),\nonumber\\
\label{firstO}
\epsilon u_y^{(1)}(r,\theta;v)\!&=&\!\frac{K_I\sqrt{r}}{4\mu\sqrt{2\pi}}\Omega_y(\theta;v)\!-\!\frac{Tr\sin\theta}{6\mu}\!+\!\C O(r^{3/2}).
\end{eqnarray}
Here $K_I$ is the Mode I ``stress intensity factor'' and $T$ is a
constant known as the ``T-stress''. Note that these parameters
cannot be determined by the asymptotic analysis as they depend on
the {\em global} crack problem. $\B \Omega(\theta;v)$ is a known
universal function \cite{98Fre,supplementary}. $\epsilon$ in
Eq.(\ref{expansion}) can be now defined explicitly as
$\epsilon\!\equiv\!K_I/[4\mu \sqrt{2\pi \ell(v)}]$, where $\ell(v)$
is a velocity-dependent length-scale. $\ell(v)$ defines the scale
where only the order $\epsilon$ and $\epsilon^2$ problems are
relevant. It is a {\em dynamic} length-scale that marks the onset of
deviations from a linear elastic constitutive behavior.

The solution of the order $\epsilon$ equation, i.e. Eqs.
(\ref{firstO}), can be now used to derive the second order problem
in $\epsilon$. The form of the second order problem for an
incompressible material is
\begin{equation}
\mu\nabla^2{\B u^{(2)}}+3\mu\nabla(\nabla\cdot{\B u^{(2)}})+
\frac{\mu\ell\B g(\theta;v)}{r^2}=\rho\ddot{\B u}^{(2)}\ .
\label{secondO}
\end{equation}
The boundary conditions at $\theta\!=\!\pm\pi$ become
\begin{equation}
\label{BC2} r^{-1}\pa_\theta u_x^{(2)}+\pa_r
u_y^{(2)}=4r^{-1}\pa_\theta u_y^{(2)}+2\pa_r
u_x^{(2)}+\frac{\kappa(v)\ell}{r}=0,
\end{equation}
where contributions proportional to $T$ were neglected. Here $\B g(\theta;v)$ and $\kappa(v)$ are known functions, see \cite{supplementary} and below.

The problem posed by Eqs. (\ref{secondO})-(\ref{BC2}) has the structure of an effective
LEFM problem with a body force $\propto\!r^{-2}$ and a crack face
force $\propto\!r^{-1}$. Note that Eqs. (\ref{secondO})-(\ref{BC2})
are valid in the range $\sim\!\ell(v)$, where $\epsilon^2$ is
non-negligible with respect to $\epsilon$, but higher order
contributions are negligible. Since one cannot extrapolate the
equations to smaller length-scales, no real divergent behavior in
the $r\!\to\!0$ limit is implied. We stress that the structure of
this problem is universal. Only $\B g(\theta;v)$ and $\kappa(v)$
depend on the second order elastic constants resulting from
expanding a {\em general} $U(\B F)$ to second order in $\epsilon$. For example, the $\propto\!r^{-2}$
effective body-force in Eq. (\ref{secondO}) results from terms of
the form $\pa(\pa u^{(1)} \pa u^{(1)})$, which are generic quadratic
nonlinearities.

We now focus on solving Eq. (\ref{secondO}) with the boundary
conditions of Eqs. (\ref{BC2}) for the explicit $\B g(\theta;v)$ and
$\kappa(v)$ derived from Eq. (\ref{NH}) \cite{supplementary}. Our
strategy is to look for a particular solution of the inhomogeneous
Eq. (\ref{secondO}) {\em without} satisfying the boundary conditions
of Eqs. (\ref{BC2}) and then to add to it a solution of the
corresponding homogeneous equation that makes the overall solution
consistent with the boundary conditions. We find that the
inhomogeneous solution, $\B \Upsilon(\theta;v)$, is r-independent.
The homogeneous solution is obtained using a standard approach
\cite{98Fre} by noting that the second boundary condition of Eqs.
(\ref{BC2}) requires that its first spatial derivative scales as
$r^{-1}$. The solution of the second order problem for Mode I
symmetry is
\begin{widetext}
\begin{eqnarray}
\label{solution}
\epsilon^2u_x^{(2)}(r,\theta;v)&=&\left(\frac{K_I}{4\mu \sqrt{2\pi}}\right)^2\left[A\log{r}+\frac{A}{2}\log{\left(1-\frac{v^2\sin^2\theta}{c_d^2} \right)}+B\alpha_s\log{r}+\frac{B \alpha_s}{2}\log{\left(1-\frac{v^2\sin^2\theta}{c_s^2} \right)}+\Upsilon_x(\theta;v)\right],\nonumber\\
\epsilon^2u_y^{(2)}(r,\theta;v)&=&\left(\frac{K_I}{4\mu
\sqrt{2\pi}}\right)^2\Big[-A\alpha_d\theta_d-B\theta_s+\Upsilon_y(\theta;v)\Big],
\quad
\tan{\theta_{d,s}}=\alpha_{d,s}\tan{\theta},\quad\alpha^2_{d,s}\equiv1-v^2/c_{d,s}^2
\ ,
\end{eqnarray}
\end{widetext}
where $A\!=\![2\alpha_s
B-4\pa_\theta\Upsilon_y(\pi;v)-\kappa(v)]/(2- 4\alpha_d^2)$ (cf. Eq.
(\ref{BC2})) and $c_d$ is the dilatational wave speed.
$c_d\!=\!2c_s$ in an incompressible material under plane stress
conditions.

The analytic form of $\B \Upsilon(\theta;v)$ depends mainly on $\B g(\theta;v)$. The latter can be represented as
\begin{equation}
g_x(\theta;v)\!\simeq\!\sum_{n=1}^{N(v)}\! a_n(v) \cos(n\theta),~~ g_y(\theta;v)\!\simeq\!\sum_{n=1}^{N(v)}\! b_n(v) \sin(n\theta).
\end{equation}
For $v\!=\!0$ we have $N(0)\!=\!3$ and the representation is {\em exact}, while for higher velocities it provides analytic approximations with
whatever accuracy needed. For $v\!\simeq\!0.8c_s$ only seven terms provide a representation that can be regarded exact for any practical purpose \cite{supplementary}.
$\B \Upsilon(\theta;v)$ is then obtained in the form
\begin{equation}
\Upsilon_x(\theta;v)\!\simeq\!\sum_{n=1}^{N(v)}\!\! c_n(v) \cos(n\theta),~~ \Upsilon_y(\theta;v)\!\simeq\!\sum_{n=1}^{N(v)}\! \!d_n(v) \sin(n\theta),
\end{equation}
where the unknown coefficients are determined by solving a linear
set of equations \cite{supplementary}. A striking feature of Eqs.
(\ref{solution}) is that they lead to strain contributions that vary
as $r^{-1}$, which are ``more-singular'' than the $r^{-1/2}$ strains
predicted by LEFM.

We now show that the second order solution of Eqs.
(\ref{solution}) entirely resolves the discrepancies raised by
trying to interpret the experimental data of \cite{companion} in
the framework of LEFM. The complete second order asymptotic
solution, Eqs. (\ref{expansion}),
(\ref{firstO}) and (\ref{solution}), contains three parameters
($K_I$, $T$ and $B$) that cannot be determined from the
asymptotic solution and therefore must be extracted from the
experimental data.

\begin{figure}
\centering \epsfig{width=.49\textwidth ,file=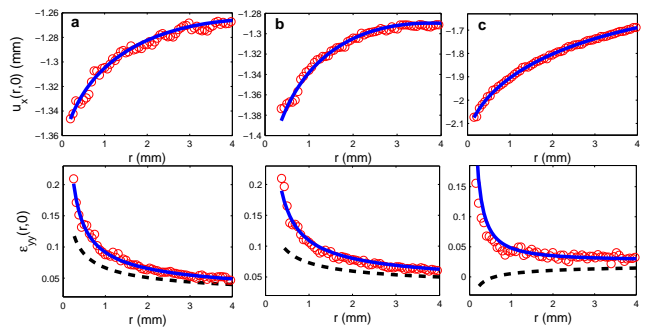}
\caption{(Color online) Top: Measured $u_x(r,0)$ (circles) fitted to
the x component of Eq. (\ref{expansion}) (solid line) for (a)
$v\!=\!0.20c_s$ with $K_I\!=\!1070$Pa$\sqrt{m}$, $T\!=\!-3150$Pa and
$B\!=\!18$. (b) $v\!=\!0.53c_s$ with $K_I\!=\!1250$Pa$\sqrt{m}$,
$T\!=\!-6200$Pa and $B\!=\!7.3$ and (c) $v\!=\!0.78c_s$ with
$K_I\!=980$Pa$\sqrt{m}$, $T\!=\!-6900$Pa and $B\!=\!26$. Bottom:
corresponding measurements of $\varepsilon_{yy}(r,0)\!=\!\pa_y
u_y(r,0)$ (circles) compared to the theoretical nonlinear solution
(cf. Eq. (\ref{expansion})) {\em with no adjustable parameters}
(solid lines); $K_I$, $T$ and $B$ are taken from the fit of
$u_x(r,0)$. (dashed lines) LEFM predictions (analysis as in
\cite{companion}) were added for comparison.
}\label{measuredstrains}
\end{figure}

These parameters were chosen such that Eqs. (\ref{expansion}),
(\ref{firstO}) and (\ref{solution}) properly describe the measured
$u_x(r,0)$. Examples for $v/c_s\!=\!0.20$, $0.53$ and $0.78$ are
provided in Fig. \ref{measuredstrains} (top). With $K_I$, $T$ and
$B$ at hand, we can now test the theory's predictions for
$\varepsilon_{yy}(r,0)$ with {\em no adjustable free
parameters}.  The corresponding results are compared with both
the measured data \cite{companion} and LEFM predictions in Fig.
\ref{measuredstrains} (bottom). In general, the agreement with the
experimental data is excellent. These results demonstrate the
importance of the predicted $r^{-1}$ strain terms near the crack
tip. $\ell$ is estimated as the scale where the largest strain
component reaches values of $0.10\!-\!0.15$. For the data presented
in Fig. \ref{measuredstrains}a,
$\varepsilon_{yy}\!>\!\varepsilon_{xx}$, where
$\varepsilon_{xx}\!=\!\pa_x u_x$ is obtained by differentiating
$u_x$. Thus, $\ell$ can be read off of the bottom panel to be
$\sim\!0.5\!-\!1$mm. Similar estimates can be obtained for every
$v$, though not always does $\varepsilon_{yy}\!>\!\varepsilon_{xx}$,
e.g. Fig. \ref{measuredstrains}c.

For $v\!=\!0.53c_s$ (Fig. \ref{measuredstrains}b) the theory still
agrees well with the measurements, although some deviations near the
tip are observed. These deviations signal that higher order
corrections may be needed, though second order nonlinearities still
seem to provide the dominant correction to LEFM. For higher
velocities, it is not clear, {\em a-priori}, that second order
nonlinearities are sufficient to describe the data. In fact, the
strain component $\varepsilon_{xx}(r,0)$ for $v\!=\!0.78c_s$ reaches
a value of $\sim\!0.35$ in Fig. \ref{measuredstrains}c, suggesting
that higher order nonlinearities may be important. Nevertheless, the
second order theory avoids a fundamental failure of LEFM; at high
velocities ($v\!>\!0.73c_s$ for an incompressible material) LEFM
predicts (dashed line in Fig. \ref{measuredstrains}c) that the
contribution proportional to $K_I$ in $\varepsilon_{yy}(r,0)$
(derived from Eqs. (\ref{firstO})) becomes {\em negative}. This
implies that $\varepsilon_{yy}(r,0)$ {\em decreases} as the crack
tip is approached and becomes {\em compressive}. This is surprising, as material points straddling $y\!=\!0$ must be
separated from one another to precipitate fracture. Thus, the second
order nonlinear solution (solid line), though applied beyond its
range of validity, already induces a qualitative change in the character of the strain. This is a striking manifestation
of the breakdown of LEFM, demonstrating that elastic nonlinearities
are generally unavoidable, especially as high crack velocities are
reached. The results of Figs. \ref{measuredstrains}a-c both provide
compelling evidence in favor of the developed theory and highlight
inherent limitations of LEFM. We note that $\ell(v)$ increases with
increasing $v$, reaching values in the mm-scale at very high $v$.

Our results indicate that the widely accepted assumption of
``K-dominance" of LEFM, i.e. that there is always a region where the
$r^{-1/2}$ strain term dominates all other contributions, is
violated here. The results presented in Fig. \ref{measuredstrains}
explicitly demonstrate that quadratic nonlinearities become
important in the same region where a non-negligible $T$-stress
exists. As elastic nonlinearities intervene before the $r^{-1/2}$
term dominates the strain fields, the contributions of {\em both} of
these terms must be taken into account as one approaches the crack
tip. Since values of the $T$-stress and of $B$ are system specific, this
observation is valid for the specific experimental system under
study. They do indicate that the assumption of ``K-dominance" is not
always valid.

An additional puzzle raised in \cite{companion} was that although
the form of both $u_x(r,0)$ and the Crack Tip Opening Displacement
(CTOD) agreed with LEFM, the respective derived values of $K_I$
differed by about 20\%, cf. Fig. 3a in \cite{companion}. This puzzle
is resolved by the theory as follows. The form of the CTOD is given
by $\phi_y(r,\pm\pi)$ as a function of the distance,
$\phi_x(r,\pi)$, from the crack tip in the moving (laboratory)
frame. Substituting $\theta\!=\!\pi$ into Eqs. (\ref{expansion}),
(\ref{firstO}) and (\ref{solution}), the nonlinear theory predicts
that the CTOD remains parabolic, where the $log(r)$ term in
$\phi_x(r,\pi)$ is negligible compared to $r$. This occurs at the
{\em same} scale $\ell(v)$ at which nonlinear corrections are
essential to describe the strain at $\theta\!=\!0$, cf. Fig. 1.
Quantitatively, the parabolic CTOD can be described with $K_I$
values that differ from those describing $u_x(r,0)$ by only a few
percent with the {\em same} values of $T$ and $B$ (cf. Fig.
\ref{measuredstrains}). This small $K_I$ variation is possibly
related to sub-leading nonlinear corrections associated with the
$T$-stress and will be addressed elsewhere.

\begin{figure}
\centering \epsfig{width=.48\textwidth ,file=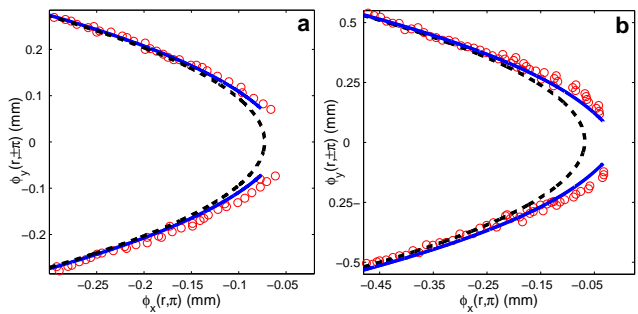}
\caption{(Color online) Measured crack tip profiles
($\phi_y(r,\pm\pi)$ vs. $\phi_x(r,\pi)$) (circles).  Shown are
the parabolic LEFM best fit (dashed line) and the profiles
predicted by  the second order nonlinear corrections (solid line).
(a) $v\!=\!0.2c_s$ and (b) $v\!=\!0.53c_s$. $T$ and $B$ are as in
Fig. \ref{measuredstrains}. In contrast to the $\sim\!20\%$
discrepancy in values of $K_I$ obtained in \cite{companion}, the
respective values $K_I\!=\!1170$Pa$\sqrt{m}$ and
$K_I\!=\!1300$Pa$\sqrt{m}$ correspond to within $9\%$ and $4\%$,
respectively, of $K_I$ obtained from $u_x(r,0)$ using the
nonlinear theory, cf. Fig. \ref{measuredstrains}.}\label{parabolas}
\end{figure}

Let us now consider the CTOD in the near vicinity of the crack tip,
i.e. when $r$ is further reduced. Eqs. (\ref{solution}) predict the
existence of $\log$-terms in $\phi_x(r,\theta)$. These terms, which
are negligible at $\theta\!=\!\pi$ on a scale $\ell(v)$, must become
noticeable at smaller scales. Although this region is formally
beyond the range of validity of the expansion of Eq.
(\ref{expansion}), we would still expect the existence of a CTOD
contribution proportional to $\log{r}$ to be observable. We test
this prediction in Fig. \ref{parabolas} by comparing the measured
small-scale CTOD to both the parabolic LEFM form and the second
order nonlinear solution {\em with no adjustable parameters}. We
find that these  $\log$-terms, whose coefficients were determined at
a scale $\ell(v)$, capture the initial deviation from the parabolic
CTOD at $\theta\!=\!\pm\pi$ to a surprising degree of accuracy. This
result lends further independent support to the validity of Eqs.
(\ref{solution}).

In summary, we have shown that the second order solution presented
in Eqs. (\ref{solution}) resolves in a self-consistent way all of
the puzzles that were highlighted in \cite{companion}. This solution
is universal in the sense that its generic properties are
independent of geometry, loading conditions and material parameters.
We would entirely expect that {\em any} material subjected to the
enormous deformations that surround the tip of a crack must
experience {\em at least} quadratic elastic nonlinearities, prior to
the onset of the irreversible deformation that leads to failure. Our
results show that these deformations, which are the vehicle for
transmitting breaking stresses to crack tips, must be significantly
different from the LEFM description, especially  at high
$v$.

One may ask why we should not consider still higher order elastic
nonlinearities. We surmise that quadratic elastic nonlinearities may
be special, as they mark the emergence of a dynamic length-scale
$\ell(v)$ that characterizes a region where material properties -
like local wave speeds, local response times and anisotropy - become
{\em deformation dependent}. This line of thought seems consistent
with the observations of Refs. \cite{Buehler03_06}. As supporting
evidence for this view, we note that the geometry-independent
wave-length of crack path oscillations discussed in
\cite{07LBDF,07BP} seems to correlate with the mm-scale $\ell(v)$ at
high $v$. Therefore, our results may have implications for
understanding crack tip instabilities.

{\bf Acknowledgements} This research was supported by grant 57/07 of
the Israel Science Foundation. E. B. acknowledges support from the
Horowitz Center for Complexity Science and the Lady Davis Trust.

\end{document}